\documentclass[acmtog]{acmart}

\AtBeginDocument{%
  \providecommand\BibTeX{{%
    \normalfont B\kern-0.5em{\scshape i\kern-0.25em b}\kern-0.8em\TeX}}}

\setcopyright{rightsretained}
\copyrightyear{}
\acmYear{0}
\acmDOI{000}

\acmJournal{TOG}
\acmVolume{0}
\acmNumber{0}
\acmArticle{0}
\acmMonth{0}

\renewcommand\footnotetextcopyrightpermission[1]{}
\begin{document}

\title{RadVR: A 6DOF Virtual Reality Daylighting Analysis Tool}

\author{Mohammad Keshavarzi}
\orcid{0000-0003-2881-165X}
\email{mkeshavarzi@berkeley.edu}
\affiliation{%
  \institution{Department of Architecture, XR Lab, University of California, Berkeley}
  \city{Berkeley}
  \state{California}
    \country{USA}
  \postcode{94720}
}
\author{Luisa Caldas }
\email{lcaldas@berkeley.edu}
\affiliation{%
  \institution{Department of Architecture, XR Lab, University of California, Berkeley}
  \city{Berkeley}
  \state{California}
    \country{USA}
  \postcode{94720}
}

\author{Luis Santos }
\email{lsantos2@kent.edu}
\affiliation{%
  \institution{College of Architecture and Environmental Design, Kent State University}
  \city{Kent}
  \state{Ohio}
    \country{USA}
}


\begin{abstract}
This work introduces RadVR, a virtual reality tool for daylighting analysis that simultaneously combines qualitative assessments through immersive real-time renderings with quantitative physically correct daylighting simulations in a 6DOF virtual environment. By taking a 3D building model with material properties as input, RadVR allows users to (1) perform physically-based daylighting simulations via Radiance, (2) study sunlight in different hours-of-the-year, (3) interact with a 9-point-in-time matrix for the most representative times of the year, and (4) visualize, compare, and analyze daylighting simulation results. With an end-to-end workflow, RadVR integrates with 3D modeling software that is commonly used by building designers. Additionally, by conducting user experiments we compare the proposed system with DIVA for Rhino, a Radiance-based tool that uses conventional 2D-displays. The results show that RadVR  can provide promising assistance in spatial understanding tasks, navigation, and sun position analysis in virtual reality. 
\end{abstract}



\maketitle
\thispagestyle{empty}

\section{Introduction}
In the past decade, with the widespread adoption of consumer-friendly and affordable hardware, Virtual Reality (VR) has gained a larger role in the Architecture, Engineering, and Construction (AEC) community. Studies suggest that immersive environments - which are not limited to visual immersion - enable a better spatial understanding when compared to 2D or non-immersive 3D representations \cite{Schnabel_Kvan_2003,paes2017immersive} enhance collaboration and team engagement among stakeholders \cite{bassanino2010impact,berg2017industry,Fernando2013} for designers and researchers to conduct virtual building occupant studies \cite{kuliga2015virtual, adi2014using, Heydarian_Carneiro_Gerber_Becerik-Gerber_2015, Heydarian_Pantazis_Carneiro_Gerber_Becerik-Gerber_2015}.  In such context, immersive visualization can be integrated in the design process as a tool that supports decision-making tasks, design modifications, and provide information on their resulting impact through building performance simulation \cite{caldas2019design}. 

In performance-oriented sustainable design, daylighting is considered a major driver behind energy consumption and occupant well-being particularly in large commercial buildings. The U.S. Energy Information Administration (EIA) \cite{center2020annual} estimates residential and commercial sectors combined use about 8\% of total electricity consumption for lighting. In this domain, simulation tools incorporate visual and photometric metrics to study occupant experience and visual comfort. However, daylight assessment of a design or a building is not limited to quantitative metrics, and other factors such as geometry and visual factors play an extensive role in this process. Such properties have been a cornerstone for daylight research, with previous studies proposing tools for objective-driven daylight form-finding \cite{Caldas_Santos_2016,caldas2008generation,Caldas_Norford_2002}, merging the spatial and visual qualities of which daylight can offer, with numeric goal-oriented generative design strategies.

Although virtual immersive environments have been widely incorporated in various design and engineering tasks, some important limitations of the current state of the technology can result in critical drawbacks in  decision making processes in design,  particularly if the design process involves daylighting assessments and analysis. In daylighting design the user highly depends on visual feedback and rendered information, and therefore, graphical and display limitations can path the way to misleading visual representations provided by the system. As a result, it is vital that daylight design VR tools identify and address this limitation.  For daylight simulation and graphics renderings, ray-tracing has been a widely accepted method in computer graphics and radiometric simulations. Following the rendering equation introduced by Kajiya \cite{Kajiya_1986}, many raytracing methods have been developed since then to simulate light behavior and optical effects. Tools for simulating daylighting performance metrics such as Radiance and Velux Daylight Visualizer take advantage of such ray-tracing techniques and have been  validated through numerous studies. As a result, these tools are broadly used in building performance design and analysis, assisting architects and building engineers to evaluate daylight behavior in different phases of the design process.

\begin{figure*}
  \centering
  \includegraphics[width=1.9\columnwidth]{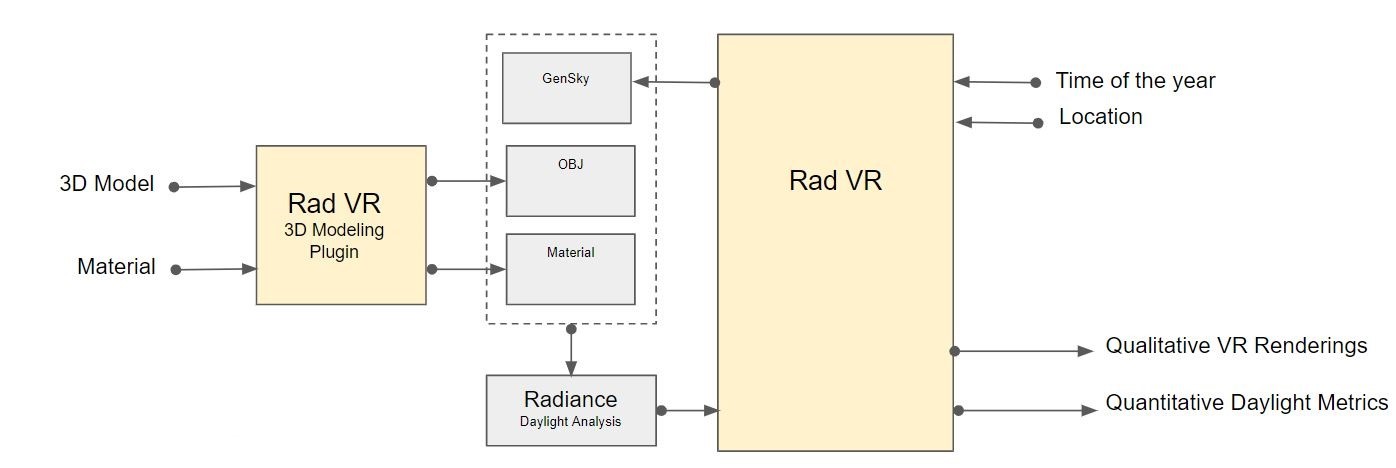}
  \caption{Workflow of RadVR- the system takes a 3D model with material properties as input. By incoperating Radiance as its calculation engine, RadVR simultaneously encompasses the qualitative immersive presence of VR and quantitative physically correct daylighting calculations of Radiance by overlaying simulation data to spatial immersive experiences.}~\label{fig:flowchart}
\end{figure*}

However, implementing real-time raytracing \cite{cook1984distributed} methods in virtual environments is challenging due to current graphic processing limitations of conventional hardware. This has resulted in the inability to produce physically correct renderings in high-frequency rates using such systems. In order to experience 6DOF and avoid user discomfort within immersive environment, rendered information displayed in Head-mounted Displays (HMD) is required to update in a framerate of least 90Hz to match the pose and field of view on the user. Rendering in such high frequencies requires lots of graphical computation power, which current conventional hardware GPU’s is  unable to provide. In addition to updating pose estimation, the wide field of view experienced in virtual environments requires high-resolution output, adding complexity and computation load to the rendering process. While there is ongoing research focused on producing high-frequency physically correct rendering which can potentially overcome the limitation of misleading renderings for daylighting designers in VR, the ability to dynamically evaluate and analyze the modeled spaces using quantitative metrics can still be considered an open challenge in the daylighting-design process.

Therefore, current game engines, which are the main development platform for virtual reality applications, take advantage of biased rendering methods to achieve faster scene processing \cite{Gregory_2018}. Such methods limit the number of ray samples and their corresponding bouncing count from the camera to the light sources (or vice-versa) resulting in unrealistic lighting representation of the target environment. However, as light bounces are limited in such methods and cannot illustrate accurate illuminance values of a given viewpoint, ambient lighting of surfaces is not achieved and is mainly limited to shadow and occlusion calculation of a scene. Many methods have been introduced to overcome this limitation \cite{williams1983pyramidal,crow1984summed,Segal_Foran_1992}, for example applying visual illusion techniques or pre-rendered texture maps, in which pre-baked light textures are efficiently mapped to the corresponding geometry in the scene, decreasing the real-time rendering load of the model. However, in applications that lighting conditions change, such methods cannot be implemented. Additionally, display limitations in current HMD systems can also decrease the required fidelity for daylighting design and decision making. Although there are several studies that propose prototypes of high dimensional range monitors, current consumer HMD hardware such as the Oculus Rift and the HTC Vive have a maximum brightness of no more than 190.5 cd/m$^2$ \cite{mehrfard2019comparative}. Therefore, relying on visual outputs of current VR rendering pipelines might be misleading and counterproductive for daylighting design processes. Hence, it is important to inform the user of possible errors and photometric mismatches through an extended visualization medium, that allows the user to compare and analyze rendered information with quantitative values in the form of common daylighting metrics.

Radiance \cite{Ward_J._1994} is the most widely used, validated, physically-based raytracing program in lighting and daylight simulation of buildings \cite{ochoa2012state, reinhart2011daylight}. Although it was one of the first backward raytracing programs developed for the light and building analysis, the incremental improvement and extension of Radiance’s raytracing capabilities to both bi-directional \cite{mcneil2013validation, geisler2016validation} and forward raytracing \cite{schregle2004daylight, grobe2019photon, grobe2019photonimage} led to its being widely regarded as the “golden standard” for lighting simulation \cite{santos2018comparison}. As a result, Radiance has been used in several inter-program comparisons and validation studies \cite{bellia2015impact, jones2017experimental,reinhart2011daylight,reinhart2009experimental}. 
 Despite the recent advancements to integrate Radiance in current digital building design workflows, there are few works that use Radiance as an ancillary analysis tool in immersive environments \cite{wasilewski2017, Jones_2017}. 

Hence, to address the challenge of using accurate quantitative daylight information in immersive environments, this work proposes an end-to-end 6DOF virtual reality tool, RadVR, that uses Radiance as its calculation engine. RadVR simultaneously encompasses the qualitative immersive presence of VR and quantitative physically correct daylighting calculations of Radiance by overlaying simulation data to spatial immersive experiences. The simulation accuracy can be customized by the user, from fast direct light analysis to progressively accurate daylighting simulations with higher detailed resolution. With an end-to-end system architecture, RadVR integrates 3D modeling software that uses conventional 2D GUI environments such as Rhino3D and provides an immersive virtual reality framework for the designer to simulate and explore various daylighting strategies. By establishing a bi-directional data pipeline between the virtual experience and Radiance, daylighting analysis can be practiced in earlier stages of design, without the need of transferring models back and forth between platforms. In our proposed methodology, overlaying quantitative calculations of various daylighting metrics within the rendered virtual space would provide additional informative value to the design process. In addition, providing the user with time-navigation and geometric manipulation tools, which are specifically developed for daylighting-based design scenarios, can further facilitate the analysis process in VR.

\thispagestyle{empty}

\section{Background}

\subsection{Radiance and Daylighting Simulation}

Radiance \cite{Ward_J._1994} is an open-source simulation engine that uses text-based input/output files, and it does not provide a Graphical User Interface (GUI). At the beginning of its development, Radiance was unable to interface with digital design tools such as Computer-Aided Design (CAD) and Building Information Modeling (BIM) programs. This forced designers and daylight analysts to use Radiance 3D modeling operations to describe a scene for simulation. Nevertheless, in the mid-1990s, Ward \cite{ward_obj2rad} proposed the obj2rad Radiance subroutine to facilitate the export of the geometry produced by a CAD and BIM tool to Radiance. Since then, several researchers and software developers have proposed several Radiance-based tools that promote the integration of design tools with the simulation engine.  

ADELINE \cite{christoffersen1994adeline, erhorn1994documentation}, was one of the first tools to integrate Radiance with a CAD tool. A small built-in CAD program, the Scribe-Modeler, provided ADELINE CAD and 3D modeling capabilities. However, ADELINE and Scribe-Modeler had several constraints regarding modeling operations and the geometric complexity of the building models, e.g., ADELINE only supported a limited number of mesh faces in a given model. ECOTECT \cite{roberts2001ecotect} was an analysis program for both thermal and daylight analysis of buildings that used Radiance to complement its limited abilities in accurately predicting daylight levels in buildings \cite{vangimalla2011validation}. The software facilitated the use of Radiance through a sophisticated GUI and became particularly popular in the early 2000s. However, the software is no longer developed and distributed. DAYSIM \cite{reinhart2001validation}, SPOT \cite{rogers2006daylighting}, and COMFEN \cite{hitchcock2008comfen} are Radiance-based tools that also emerged at the turning of the century. Although DAYSIM and SPOT extend Radiance calculation capabilities to Climate-Based Daylighting Modeling (CBDM), they do not provide a user-friendly GUI. Thus, the standalone version of these tools is seldom used by architects in their design workflows. Regarding COMFEN, its modeling and simulation abilities are extremely limited because the tool was specifically developed for initial analyses that focus on glazing selection and the design of simplified static shading systems.

DIVA for Rhino \cite{jakubiec2011diva} is a daylight and thermal analysis tool that largely contributed to the integration of Radiance-based analysis in building design workflows. DIVA is fully integrated with Rhinoceros (Rhino) \cite{mcneel2015rhinoceros}, a popular Non-Uniform Rational Basis Spline (NURBS) CAD software among architects. DIVA easily processes the geometry modeled in Rhino to Radiance and DAYSIM, and provides an easy-to-use GUI that interfaces with both simulation engines. The user can also access to DIVA through Grasshopper, a Visual Programming Language (VPL) for Rhino, to perform more advanced Radiance simulations. Ladybug+Honeybee \cite{roudsari2013ladybug} enables a complete use of DAYSIM and Radiance-based techniques, including bi-directional raytracing techniques \cite{mcneil2013validation, geisler2016validation}, through Grasshopper and Dynamo VPLs. The Ladybug+Honeybee version for Dynamo allows the use of Radiance and DAYSIM in the Revit BIM program. Autodesk Insight is another tool that also supports Cloud-based Radiance calculations for BIM design workflows.

As briefly reviewed above, there have been several efforts to develop intuitive interfaces that facilitate the use of Radiance as a lighting and daylighting simulation engine for performance-based design workflows.. Throughout the years, there have always been efforts to migrate the Radiance engine toward new interfaces by introducing intuitive design and analysis features to allow users of such systems to conduct daylighting simulation and informative analysis. Nevertheless, the integration of Radiance in performance-based design processes supported by Virtual and Augmented Reality interfaces is still in its early phases. It is important to investigate new user interfaces in Virtual Reality and Augmented Reality that are supported by state-of-the-art daylighting simulation engines such as Radiance for two reasons: (1) it is foreseeable that VR and AR will play an important role in the design of the built environment, (2) daylighting design workflows supported by AR and VR tools require the robust and reliable predictions provided by state-of-the-art simulation engines such as Radiance. This work was developed in such a research direction, by introducing a novel 6DOF virtual reality interface for the Radiance engine, for simulation, visualization, and analysis for daylighting-based design tasks.

\thispagestyle{empty}
\subsection{Virtual Reality and Design Task Performance}
 In the AEC and design community, virtual reality platforms are being gradually adopted as new mediums that can potentially enhance the sense of presence, scale, and depth of various stakeholders of building projects. Several methods have been introduced to study the relationship between spatial perception and user task performance in immersive environments. Witmer and Singer \cite{Witmer_Singer_1998} define presence as the subjective experience of being in one place or environment, even when one is physically situated in another. By developing presence questionnaires, they argue that a consistent positive relationship can be found between presence and task performance in virtual environments. Since then, similar questionnaires have been applied in several studies in AEC and related fields on VR \cite{Castronovo_Nikolic_Liu_Messner_2013, Kalisperis_Muramoto_Balakrishnan_Nikolic_Zikic_2006, keshavarzi2019affordance} with Faas et al. specifically investigating whether immersion and presence can produce better architectural design outcomes at early design stages \cite{Faas_Bao_Frey_Yang_2014}.
For skill transfer and decision making tasks, Waller et al. show that sufficient exposure to a virtual training environment has the potential to surpass a real-world training environment \cite{Waller_Hunt_Knapp_1998}. Heydarian et al. conclude users perform similarly in daily office activities (object identification, reading speed and comprehension) either in immersive virtual environments or benchmarked physical environments \cite{Heydarian_Carneiro_Gerber_Becerik-Gerber_Hayes_Wood_2015}. Moreover, other studies investigated if virtual environments enhance occupant navigations in buildings when compared to 2D screens. Some indicating a significant improvement \cite{Robertson_Czerwinski_van_Dantzich_1997, Ruddle_Payne_Jones_1999}, while others present no significant differences \cite{mizell2002comparing,Sousa_Santos_Dias_Pimentel_Baggerman_Ferreira_Silva_Madeira_2009}.

In addition to individual design task procedures, the ability to conduct productive virtual collaboration between various stakeholders in a building project is an important factor that can impact multiple stages of the design process. 
Such capabilities of immersive environments have been broadly investigated for collaborative review purposes which usually happen in the last phases of the design process. In these phases, critical evaluation and analysis can impact construction cost and speed. \cite{Eastman_Teicholz_Sacks_Liston_2011} Identifying missing elements, drawing errors and design conflicts can help to avoid unwanted costs and allocation of resources. Commercial software such as Unity Reflect, Autodesk Revit Live and IrisVR enable virtual walkthroughs and facilitate visualization of conventional 3D and BIM file formats. Some of these platforms also allow the designer to update the BIM model directly within the immersive environment or vice versa.

\subsection{Building Performance Visualization in Virtual Reality}
Multiple studies explored the visualization results for building performance simulation in VR, either by using pre-calculated data or using the VR interface to perform simulations and visualize their outputs. For example, Nytsch-Geusen et al. developed a VR simulation environment using bi-directional data exchange between Unity and Modelica/Dymola \cite{nytsch2017buildingsystems_vr}. Rysanek et al. developed a workflow for managing building information and performance data in VR with equirectangular image labeling methods \cite{rysanek2017workflow}. For augmenting data on existing buildings, Malkawi et al. developed a Human Building Interaction system that uses Augmented Reality (AR) to visualize CFD simulations \cite{Malkawi_Srinivasan_2005}. Augmented and virtual reality interfaces have also been applied for structural investigations and finite element method simulations. In \cite{Hambli_Chamekh_Bel_Salah_2006}, the authors use artificial neural networks (ANN) and approximation methods to expedite the simulation process and achieve real-time interaction in the study of complex structures. Carneiro et al. \cite{carneiro2019influencing} evaluate how spatiotemporal information visualization in VR can impact user design choices. They report participants reconsider initial choices when informed of better alternatives through data visualization overlaid in virtual environments. For performance-based generative design systems, work of \cite{keshavarzi2020v} enables users to analyze and narrow down generative design solution spaces in virtual reality. Their proposed system utilizes a hybrid workflow in which a spatial stochastic search approach is combined with a recommender system allowing users to pick desired candidates and eliminate the undesired ones iteratively in an immersive fashion.

\subsection{Daylight Analysis in Virtual Reality}

In the study of daylighting, immersive environments have been widely used as an end-user tool to study daylight performance and provide occupant feedback. In this regard, Heydarian et al. study the lighting preferences of building occupants through their control of blinds and artificial lights in a virtual environment \cite{Heydarian_Pantazis_Carneiro_Gerber_Becerik-Gerber_2015}. Rockcastle et al. used VR headsets to collect subjective evaluations of rendered daylit architectural scenes \cite{Rockcastle_Chamilothori_Andersen_2017}. Using similar settings, Chamilothori et al. experimented the effect of façade patterns on the perceptual impressions and responses of individuals to a simulated daylit space in virtual reality \cite{chamilothori2019subjective}.  Carneiro et al. \cite{carneiro2019understanding} use virtual environments to study how the time-of-day influence participants lighting choices in different window orientations. Similar to our study, they use real-time HDR rendering to visualize lighting conditions in VR. However, their study is limited to only three times of the day (9am, 1pm, 5pm) and no real-time physically-correct simulation takes place to inform users with quantitative daylighting metrics of the target virtual environments. Instead they use pre-calculated animations to visualize lighting levels. 

There is also a body of research which examines the validity of using VR to represent the impressions of a scene for various lighting conditions. Chen et al. \cite{chen2019virtual} compare participants subjective feeling towards a physical lighting environment with a virtual reality reproduction, video reproduction and photographic reproduction. They show human subjects are most satisfied with VR reproductions with a coefficient of 0.886. Chamilothori et al. \cite{chamilothori2019adequacy} used 360 degree physically-based renders visualized in a VR headset to evaluate and compare perceived pleasantness, interest, excitement, complexity, and satisfaction in daylit spaces with a real-world setting. While such approach can be heavily dependent on the tone-mapping method used, their results indicate no significant differences between the real and virtual environments on the studied evaluations. Another similar validation is seen in Abd-Alhamid et al. work \cite{abd2019developing} where they investigate subjective (luminous environment appearance,and high-order perceptions) and objective (contrast-sensitivity and colour-discrimination) visual responses in both real and VR environments of an office. They also report no significant differences of the studied parameters between the two environments and show a high level of perceptual accuracy of appearance and high-order perceptions in VR when compared to the real-world space.

Finally, for real-time quantitative simulation and visualization, Jones \cite{Jones_2017} developed Accelerad, a GPU accelerated version of Radiance for global illumination simulation for parallel multiple-bounce irradiance caching.The system allows faster renderings when compared to a CPU version of Radiance, and therefore, can facilitate higher refresh rates for VR environments. However, the VR implementation of this method currently does not provide high-frequency 6 degrees-of-freedom (6DOF) renderings, thus being limited in providing an enhanced sense of presence and of scale to the user.. Our work, in contrast, utilizes qualitative 6DOF rendering using HDR pipelines and integrates Radiance as a backbone for quantitative simulations. Such a combination can provide a smooth (high refresh rate) spatial exploration within the immersive environment while allowing the user to overlay various daylighting simulations for quantitative analysis. In addition, we attempt to introduce novel interaction modules, to facilitate the daylighting design task in VR.

\begin{figure*}
  \centering
  \includegraphics[width=1.3\columnwidth]{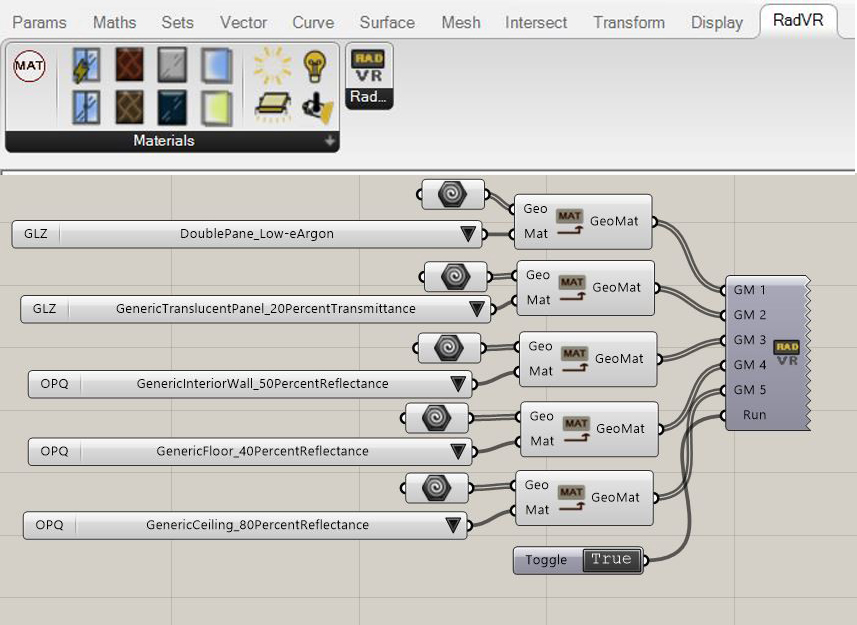}
  \caption{RadVR import plug-in for Grasshopper, a visual programming language for Rhinoceros 3D. Using the \emph{Assign Material} component, different material types (glazing, plastic, translucent, electrochromic glazing, etc.) can be assigned to geometry, and provide a semantic input for RadVR}~\label{fig:ghComp}
\end{figure*}

\section{Methodology}

\thispagestyle{empty}
\subsection{System Architecture}

Figure ~\ref{fig:flowchart} shows the workflow of RadVR’s end-to-end processing pipeline. The system takes semantic 3D geometry as input and automatically converts it to a Radiance geometry description with the corresponding material properties. When the user runs RadVR within the virtual environment, the Radiance engine runs in the background an initial simulation to prepare the primary scene within VR. This loads the entire geometry with its defined material into VR, allowing the designer to explore, simulate and review the multiple daylighting functions of the tool. From this moment on, the virtual reality software simultaneously integrates the Radiance engine in performing different simulations in a bi-directional manner, allowing the user to trigger the simulations directly in VR.

The following subsections firstly describe the core issues addressing the design of the system architecture; semantic 3D geometry input, Octree preparation, Radiance integration, and game engine implementation. Secondly, we discuss different functionalities of RadVR, simulations types, visualization, and output metrics. Finally, we describe the design approach of the user interaction of RadVR and implementation of the different modules of the system.

\subsubsection{Semantic Geometry Input}

As the daylighting performance of a building is highly dependent on material properties of the target space, the procedure of importing geometry should be intertwined with a semantic material selection to building a correct scene for daylight simulations. However, at early design stages, the material assignment for certain surfaces may not be finalized. Thus, to allow early-stage daylighting analysis the authors developed a RadVR plug-in for Grasshopper - a visual programming environment for Rhinoceros3D. With this plug-in, the user directly assigns the corresponding material to every geometry prepared to be exported, including surfaces or meshes modeled in Rhino or the result of parametric geometry produced by a Grasshopper algorithm. 

The RadVR plug-in produces the required data file for both the game engine and simulation engine (Radiance) in two separate target directories. Such method allows the user to interact with one unified input module in Rhino/Grasshopper 2D GUI before transferring to a virtual immersive environment. Moreover, the plug-in serves as a bridge between 3D modeling and parametric practices with the performance analysis in VR systems. The plug-in also provides a predefined material list in which the user can choose the material or modify its main parameters by using other Grasshopper functions in the pipeline. The plugin is not a required component of the RadVR system, but is a rather a tool to facilitate importing data to the RadVR interface. Figure \ref{fig:ghComp} shows an example of how a semantic 3D model is exported via the RadVR plugin components.

\begin{figure*}
  \centering
  \includegraphics[width=1.8\columnwidth]{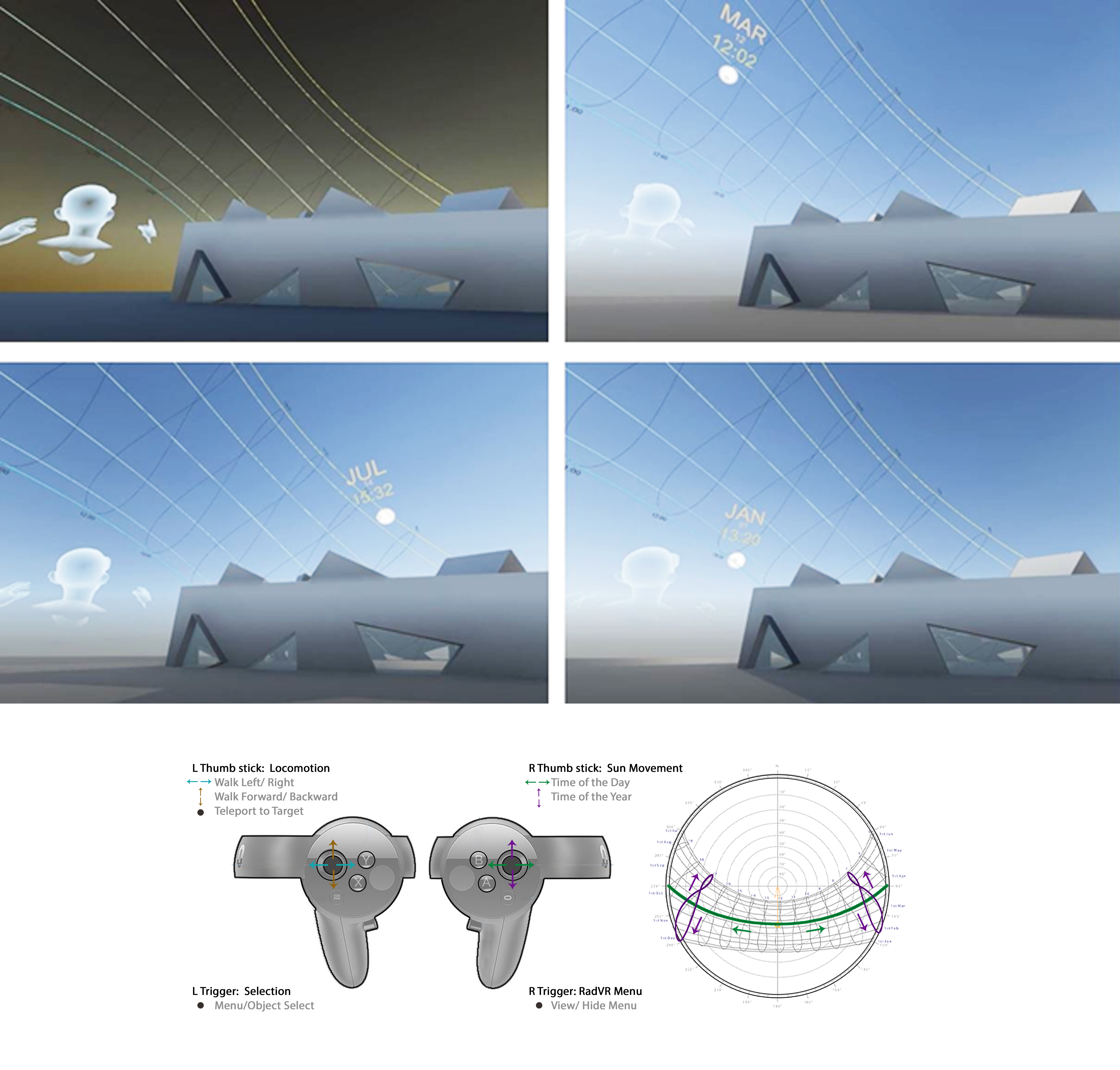}
  \caption{Changing the time of the year using virtual reality touch controllers. By pressing up/down the month of the year is modified and by pressing left/right the hour of day is modified}~\label{fig:4pics}
\end{figure*}

\subsubsection{Octree Preparation}
To prepare the input building description for daylighting simulations within VR, the system labels each instance of the geometry input to the corresponding material property and assigns a sky model to the scene first depending on simulation type, and if necessary on location, day, and hour. If the user requests a daylight factor simulation the system automatically generates an CIE overcast sky, but if the user requests an illuminance analysis it will generate a CIE clear sky model based on location, day, and hour. The sky model is stored in a dedicated text file, the labelled geometry in a Radiance file (*.rad), and another text file describes the materials optical properties. All the files are combined into a single octree. If needed, the user can modify the material file by editing or writing their own materials using dedicated Radiance shaders.    

\begin{figure*}
  \centering
  \includegraphics[width=2\columnwidth]{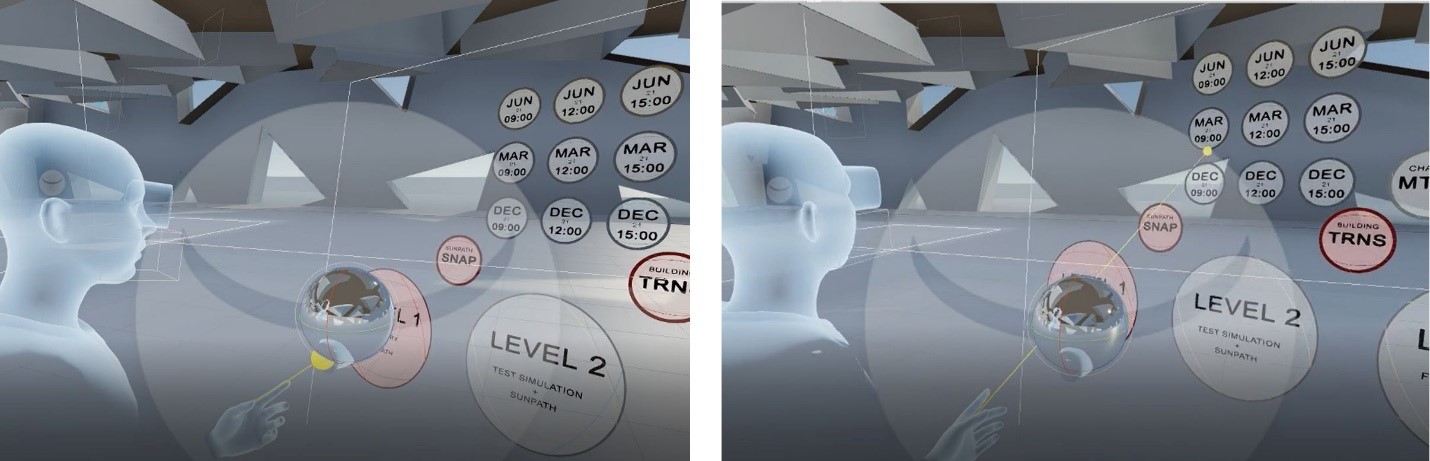}
  \caption{9 point-in-time matrices in RadVR. While choosing each date in the matrix, the sun position instantly updates to construct the corresponding shadows and daylighting effects.}~\label{fig:9point}
\end{figure*}

\thispagestyle{empty}
\subsubsection{Radiance Integration}
For raycasting-based daylight simulation, RadVR uses Radiance \cite{Ward_J._1994} as its calculation engine. Radiance is a validated daylighting simulation tool \cite{mardaljevic1995validation, reinhart2000simulation, reinhart2006development} developed by Greg Ward and it is composed by a collection of multiple console-based programs. To simulate irradiance or illuminance values at individual sensors the system uses Radiance’s subprogram \emph{rtrace}. These sensors may form a grid over a work plane, or they may represent individual view directions for pixels of an image. However, instead of calculating color values of output pixels of a scene,  \emph{rtrace} sensors can be implemented in a wide range of spatial distribution covering target locations with multiple direction in an efficient manner. This minimizes computation time by limiting ray tracing calculations to specific targets and avoid calculating images of a large size as one directional array. 

When a simulation is triggered in RadVR a designated C\# script is activated to communicate with Radiance  \emph{rtrace} through the native command console. The required input of every simulation is provided according to the virtual state of the user and time of the year of defined in GUI of RadVR. Moreover, the  \emph{rtrace} simulation runs as a background process without the user viewing the simulation console or process. Once the simulation is complete, a virtual window notifies the user of the completion and the scene would be updated with the simulation visualization. The results are stored in memory and can be later parsed and visualized if called by the user. 

\subsection{Game Engine Implementation}
As discussed in the introduction, the ability to output high frequency renderings in an efficient manner is the main objective of modern game engines. RadVR uses the Unity3D game engine and libraries for its main development platform. Like many other game engines, Unity is currently incapable of real-time raytracing for VR applications and implements a variety of biased-rendering methods to simulate global lighting. For material visualization, a library of Unity material files was manually developed using the Standard Shader. The parameters of the shaders were calibrated by the authors to visually correspond to the properties of listed materials available in Grasshopper plugin. In addition, these properties can be later modified in RadVR which would be visually updated during runtime.


\thispagestyle{empty}
\subsection{RadVR User Modules}
\subsubsection{Direct Sunlight Position Analysis}

One important aspect of daylight analysis is understanding the relationship between time, sun position, and building geometry. Hence, we implemented a module that given a building location (latitude and longitude), a day of the year and a hour of the day it correctly positions the sun for direct sunlight studies in buildings. In RadVR, an interactive 3D version of the stereographic sun path diagram is developed with calculations of the US National Oceanic and Atmospheric Administration (NOAA) Sunrise/Sunset and Solar Position Calculators. These calculations are based on equations from Astronomical Algorithms, by Jean Meeus \cite{Meeus_1998} . Each arc represents a month of the year and each analemma represents an hour of the day.

The authors implemented a C\# script to translate NOAA equations to functions that operate within the Unity3D environment. This script calculates the zenith and azimuth of the sun based on longitude, latitude and time of the year, and controls the position and rotation components of a direct light object in the VR environment. The mentioned inputs are accessible from the implemented GUI of the program, both through user interface menu options and VR controller input.

\thispagestyle{empty}

To avoid non-corresponding arcs throughout the months, the representing days of each month differ and are as follows: January 21, February 18, March 20, May 21, June 21, July 21, August 23, September 22, October 22, November 21, December 22. In addition, monthly arcs are color coded based on their season with the winter solstice (December 22 in northern hemisphere and June 21 in southern hemisphere) visualized in blue, and the summer solstice arc (June 21 in northern hemisphere and December 22 in southern hemisphere) color coded in orange. Monthly arcs in between correspond to a gradient of blue and orange based on their seasonality.

The observer location of the sun path diagram is set to the center eyepoint (mid-point between the virtual left and right eyes). The diagram moves with the user, with its center always positioned at the observer location. When the user is turning around their head or executing virtual locomotion within the immersive environment, the sunpath diagram location is updated. This feature of the software also allows users to indicate whether direct sun illumination is visible from the observer’s position throughout the year. If a section of sun-path diagram is visible through the building openings it indicates direct sunlight penetration will happen during the corresponding time at the observer’s location.

The user controls the time of day using two different input methods. The first is by using VR controllers and changing the time with joystick moves. On moving the joystick from left to right the time of the day increases on a constant day of the year and vice versa. The joystick movement results on the sun’s arc movement from sunrise to sunset. In contrast, on moving the joystick from down to up, the day of the year increases in a constant time of the day, which results moving on the corresponding analemma in the sun path diagram. The speed of the movement can be adjusted through RadVR settings, allowing users to control their preferable sun path movement for daylight analysis. Moreover, to adjust the time in hourly steps and avoid the smooth transition in minutes, authors implemented a  \emph{SnapTime} function to assist user designers in altering time of the day controls. This function also extends to the day of the year, with snaps happening on the 21st of the month only. Designed as an optional feature, which can be turned on or off using the RadVR menu, \emph{SnapTime} allows users to quickly and efficiently round the time of the year on hourly and monthly numbers for sunlight analysis. The second input method is using an immersive menu, which is loaded when the user holds the trigger. Using a raycasting function, the user can point toward different buttons and sliders and select the intended time of the day and day of the year.

When the time of day and date of the year is being changed by the user, lighting conditions and shadows are updated based on the corresponding building model. However, in many cases the user is eager to locate the position of the sun relative to the building, but due to the specific geometry of the model, the sun location is being blocked by the solid obstructions. To resolve this issue, we implemented the \emph{Transparent Model} function in the workflow, which adds a see-through effect to the model to observe the sun position relatively to the point-of-view. The \emph{Transparent Model} function replaces all solid and translucent materials with a transparent material to achieve this.

While conventional daylighting analysis tools such as DIVA also provide visualization of the sunpath diagram in their 2D interfaces, we believe our human-centric immersive approach would further facilitate the task of correlating the 3D attributes of a designed building to the 3D properties of the sunpath. In such context, our goal was to develop an easy-to-control sunpath interaction module where the user can intuitively inspect in real time the relationship between the position of the sun, which depends on time, the building, and the resulting direct light pattern while freely moving around in the building. Our approach also exceeds sunpath simulators of current BIM visualization software in VR, such as IrisVR and Unity Reflect, by rendering analemmas, generating the 9 point-in-time matrix (i.e., 9 relevant hours that cover the different equinox and solstices),, and triggering transparent material modes to allow a better understanding of the sunpath movement and it's relation to the building.

\subsubsection{The 9 point-in-time matrix}\label{sec:9Point}

In addition to the manual configuration of time, assessing a 9 point-in-time matrix is also a useful method used in daylight studies. The analysis of the morning (9:00am), noon (12:00pm) and afternoon (3:00pm) for the solstices and equinoxes is a fast way to evaluate and compare daylight patterns throughout the year. Users access the RadVR version of point in time matrix through the corresponding UI menu which contains 9 captioned buttons that represent the 9 point in times. By clicking on each button, the time would be updated in the surrounding environment, resulting in the reposition of the sun position, shadows, etc. In contrast to the conventional 9 point-in-time matrix where a single viewpoint of the building is rendered in 9 times in different times of the year to form a 3x3 matrix of the rendered viewpoints in one frame, RadVR’s 9 point-in-time is a set of nine 360 degrees 6DOF viewpoints that are individually accessed and updated through the 3x3 user interface shortcut. Therefore, evaluation of these times can be done in a much wider field of view covering all surroundings and not just one specific camera angle. This may result in a more comprehensive daylight comparison of the buildings space, as user designers can simultaneously identify geometrical properties of daylight in multiple viewpoints of the buildings. However, the limitation of not being able to view all renderings in one frame can be viewed as a drawback compared to the conventional 9 point-in-time render matrix. For an in-depth comparison of conventional 2D point-in-time matrix vs RadVR, please see results of user studies in Section \ref{sec:user}.

\thispagestyle{empty}
\subsubsection{Quantitative Simulations}
\begin{figure*}
  \centering
  \includegraphics[width=1.6\columnwidth]{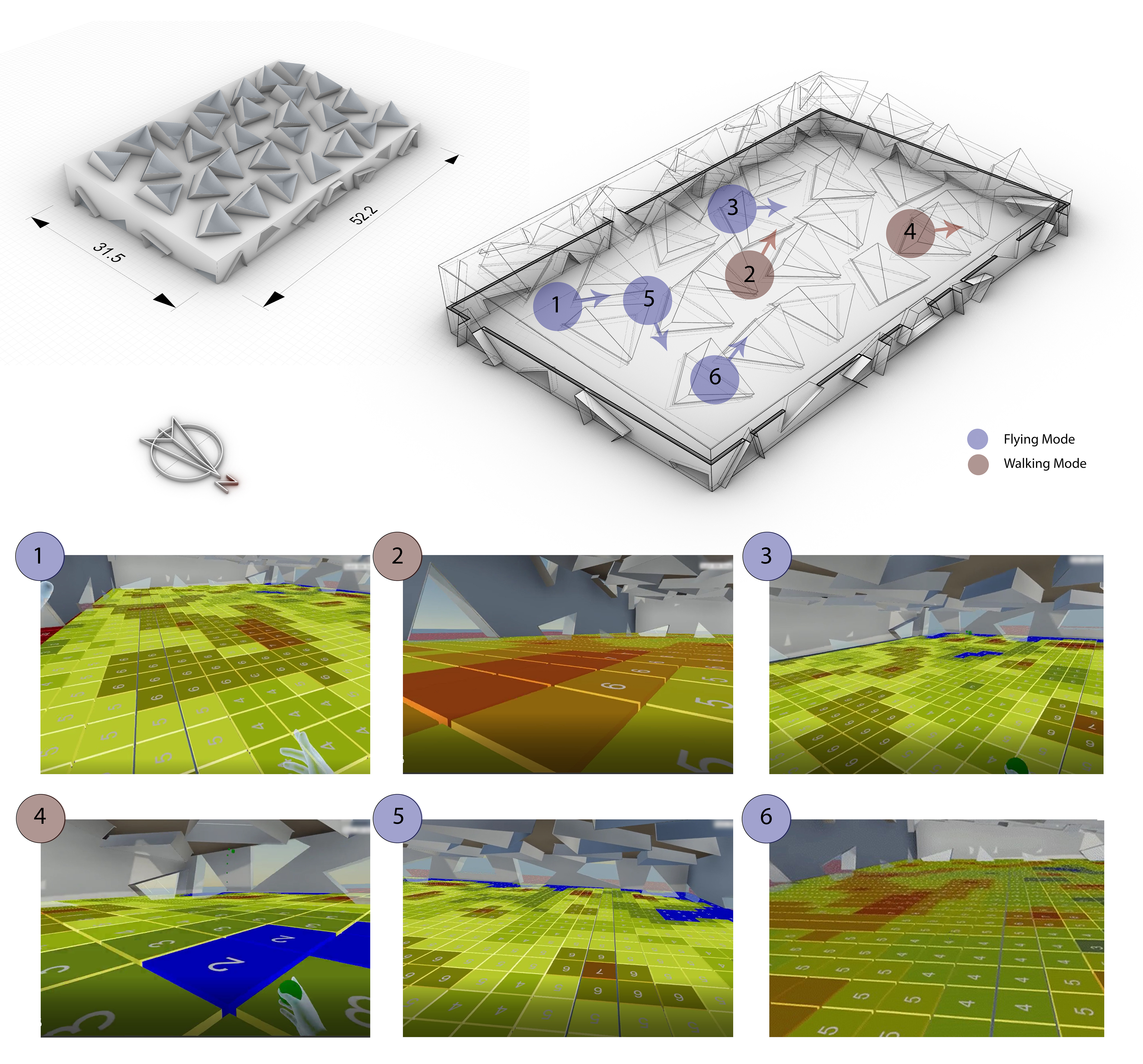}
  \caption{Example visualization of a Daylight Factor $(DF)$ simulation within RadVR. Value are plotted in the location of each sensor node. A three-color gradient palette is implemented where blue is considered as the minimum value, yellow  as the median, and red as the maximum value. The range of heatmap can be modified through RadVR menus.}~\label{fig:visualization}
\end{figure*}

One of the main design objectives of RadVR is to allow the user to spatially map the daylighting performance of the building with its geometrical properties while immersed in the virtual model of the building itself. When visualizing on-demand simulations in the surrounding immersive space, the user would be able to perceive which geometrical properties are impacting the results by visually inspecting the building with 6DOF movement (see Fig \ref{fig:visualization}). While qualitative renderings of the daylit scene are produced directly from the game engine rendering pipeline, the physically correct quantitative simulations of conventional daylighting metrics are achieved by triggering Radiance simulations through the front-end user interface of RadVR. By defining certain simulation settings such as simulation type, sensor array resolution, and ambient bounce count through user-centric interaction modules, the designer can run, visualize, compare and navigate through different types of daylighting simulations within the immersive environment of RadVR. The following explains the different components of the quantitative simulation front-end module:

A standard workflow for illuminance-based simulations, which is also used in other daylighting analysis software such as DIVA, is to define an array of planar sensors and measure illuminance in each sensor. The sensors description includes spatial position given by cartesian 3D coordinates and a vector that defines its direction. For daylighting simulation, the sun location and sky conditions define the lighting environment, therefore the time of the desired simulation and its corresponding sky model are input parameters for the simulation.

\begin{figure*}
  \centering
  \includegraphics[width=2\columnwidth]{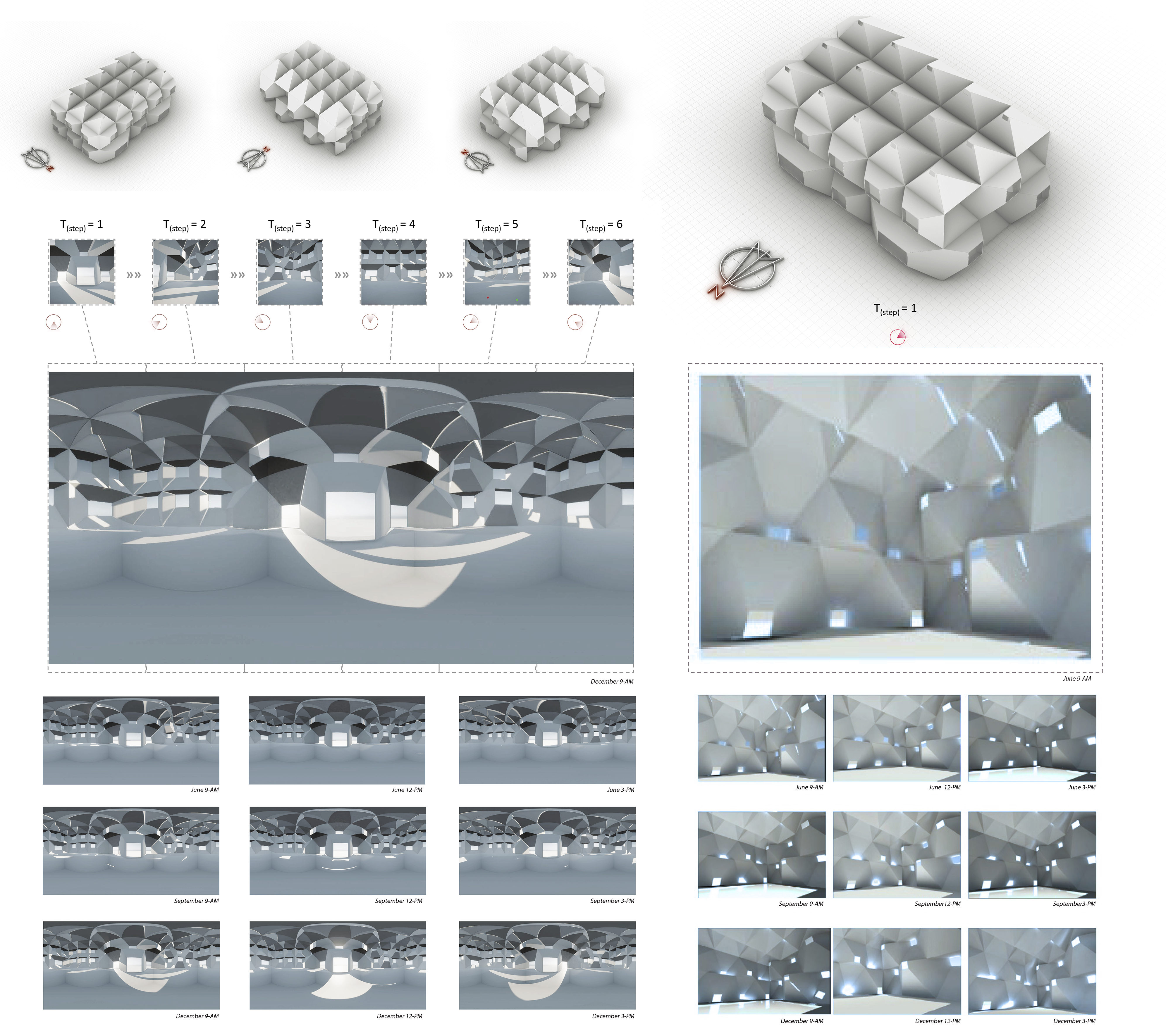}
  \caption{Comparison of 9 point-in-time matrix between the extracted screenshot panorama of RadVR (left) and Diva4Rhino (right).}~\label{fig:comp}
\end{figure*}

To construct the sensor arrays in RadVR, a floating transparent plane -  \emph{SimulationPlane} - is instantiated when the user is active in the simulation mode. This  plane follows the user within the virtual space during all types of virtual locomotion (teleporting, touchpad-walking, flying) allowing the user to place the \emph{SimulationPlane} based on their own position in space. The size and height of the  \emph{SimulationPlane} can be adjusted using corresponding sliders. This type of interaction allows the VR user to locate the simulation sensors wherever the user intends to in the virtual environment. In contrast to conventional 3D modeling software which visual feedback is inherited from birds-eye views, and orbiting transformations are considered as the main navigation interaction, immersive experiences are highly effective when designed around human-scale experiences and user-centric interactions. Therefore, instead of expecting the user to use flying locomotion navigation and accurate point selection to generate the sensor grid, the  \emph{SimulationPlane} automatically adjusts its position relative to the user position.

The distance between sensor points, can also be adjusted by the user in both X and Y directions. Such property allows the designer to control the simulation time for different testing scenarios or allocate different sensor resolutions in different locations. If studying a certain area of virtual space requires more resolution, the user can adjust the  \emph{SimulationPlane} size, height and sensor spacing, while modifying the same parameters for another simulation which can be later overlaid or visualized in the same virtual space. In addition to sensor resolution, ambient bounces of the light source rays is another determining factor in the accuracy of ray tracing simulations. While the default value of the RadVR simulations is set to 2 ambient bounces, this parameter can be modified through the corresponding UI slider to increase simulation accuracy in illuminating the scene. However, such increase results in slower calculations, a factor which the user can adjust based on the objective of each simulation.

The time and the corresponding sun location of each simulation is based on the latest time settings controlled by the user in the RadVR runtime. By using the touchpad controller to navigate the month of year and hour of the day or accessing any of the given timestamps of the 9-point-matrix, the user can modify the time of the year for the simulation setting. Moreover, longitude and latitude values can be accessed through the RadVR menu allowing comparative analysis for different locations.

The current version of RadVR offers quantitative simulations of two daylighting metrics: (a) Point-in-time Illuminance $(E)$, and (b) Daylight Factor ($DF$). In the following sections, we provide a short description of each metric and elaborate on why the mentioned metrics were prioritized in the development of the system. 

\thispagestyle{empty}
\subsubsection{Point-in-time Illuminance Simulation}

The illuminance at a point $P$ $(E_P)$ of a given surface is the ratio between the luminous flux $(\upphi)$ incident on an infinitesimal surface in the neighborhood of $P$ and the area of that surface $(A_{rec})$. $E$ measuring units are lux,in the International System of Units (SI), and foot-candles (fc) in the Imperial System of Units (IP). Illuminance basically measures how much the incident light illuminates a surface in terms of human brightness perception. The mathematical formula is \cite{carlucci2015review}:

$$E_P =  \frac{d\upphi}{dA_{rec}} \text{[lux or fc]}$$

\thispagestyle{empty}
Since $E$ measurements are local and are assessed in a point-in-time fashion, illuminance simulations have the advantage of a faster calculation process when compared to other daylighting metrics and can deliver an accurate measurement in an instantaneous moment for a specific spatial location within a given luminous environment. Such property provides RadVR users to  trigger and visualize simulations in a fast and iterative manner, while navigating and inspecting the results. However, illuminance is a point-in-time (static) metric and does not measure daylighting quality in a given period of time. Therefore, a fast and effective user workflow should be established to allow the user to iterate and compare between different time of the year to provide an efficient feedback to the design process. Such approach is followed in RadVR, in which by easily changing the time of the year with touchpad controllers or using the 9-point-in-time matrix menu the user can modify the simulation time settings in an immediate manner. The calculation of illuminance levels over a grid of sensor points involves the use of Radiance's rtrace subroutine \cite{ward1996radiance}. For more details on using rtrace and other complementary routines to calculate illuminance, the authors refer the reader to the tutorials of Compagnon \cite{compagnon1997radiance} and Jacob \cite{jacobs2014radiance}.

\thispagestyle{empty}
\subsubsection{Daylight Factor Simulation}

The Daylight Factor at a point $P$ $({DF}_P)$, is the ratio of the daylight illumination at a given point on a given plane due to the light received directly or indirectly from the sky of assumed or known luminance distribution to the illumination on a horizontal plane due to an unobstructed hemisphere of this sky. Direct sunlight is excluded for both interior or exterior values of illumination. The following expression calculates ${DF}_P$ \cite{hopkinson1966daylight}:

$${DF}_P = \frac{(E_P,obs)}{(E_P,unobs)}$$

where $(E_P,obs)$ is the horizontal illuminance at a point P due to the presence of a room that obstructs the view of the sky and $(E_P,unobs)$ is the horizontal illuminance measured at the same point P if the view of the sky is unobstructed by the room.

$DF$ was first proposed by Trotter in 1895 \cite{walsh1951early} and is based on a ratio to avoid the dependency of assessing daylight performance based in instantaneous sky conditions \cite{reinhart2006dynamic}. $DF$ assumes that the sky has a uniform luminance, therefore, it uses an overcast sky model in the simulation process. Such assumption results in fast simulation time and can be representative for an entire year. As $DF$ cannot properly represent daylight illumination conditions that differ from the overcast sky \cite{mardaljevic2009daylight} and is insensitive to building orientation \cite{kota2009historical}, its combination with illuminance point-in-time simulations are useful to a broader understanding of the daylighting performance of the space, while maintain a fairly low simulation run-time and computation process. The calculation of Daylight Factor and a given point also uses rtrace \cite{ward1996radiance}. For more details on using rtrace and other complementary routines to compute Daylight Factor, the authors refer the reader to the tutorials previously mentioned in the calculation of illuminance \cite{compagnon1997radiance,jacobs2014radiance}.

\begin{figure*}
  \centering
  \includegraphics[width=2\columnwidth]{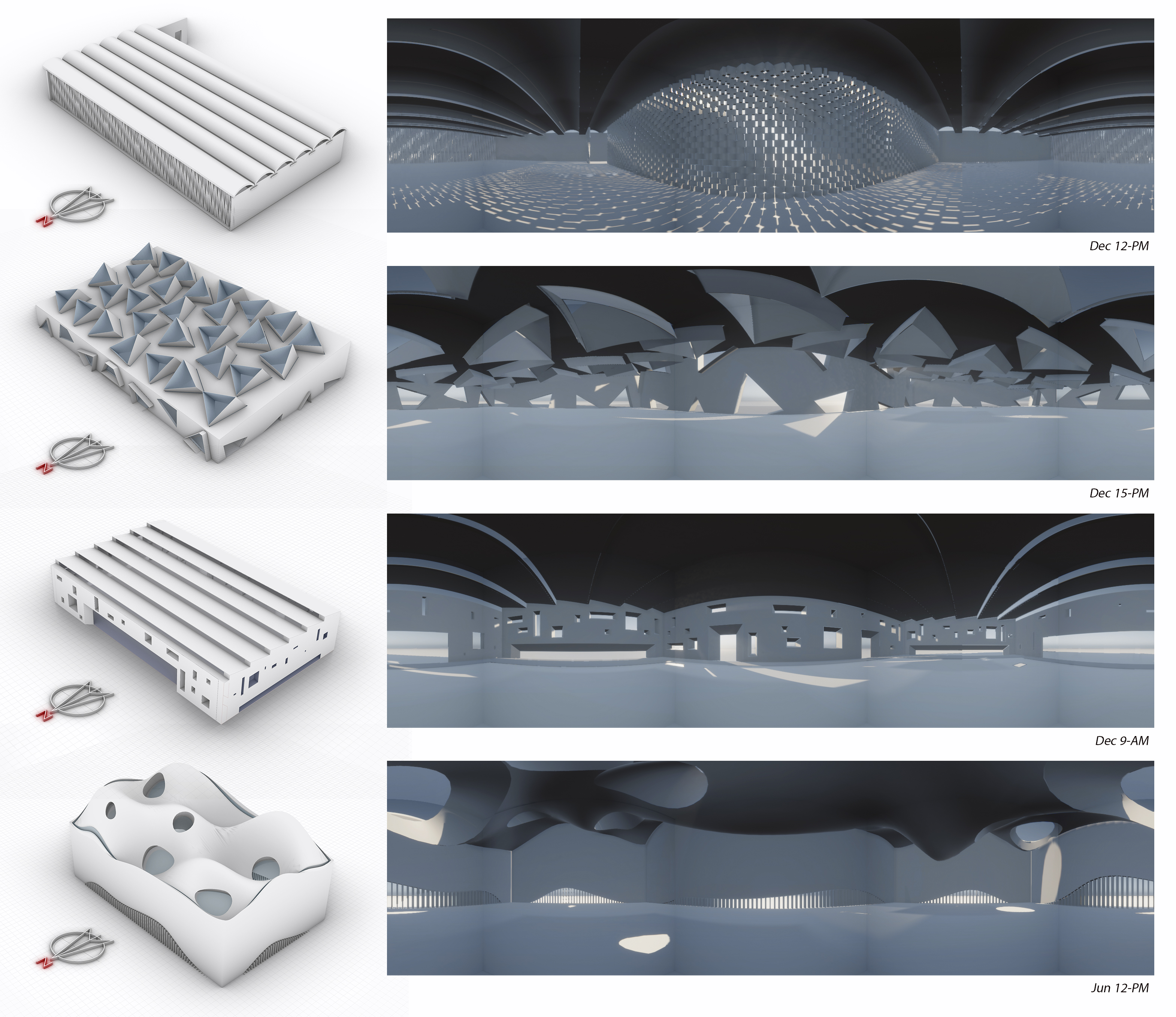}
  \caption{Interior panoramic screenshot of a single frame using the RadVR system. Renderings are executed in real-time providing a 6 degrees of freedom immersive experience.}~\label{fig:othergroup}
\end{figure*}

\thispagestyle{empty}
\subsection{Visualization of Simulation Results}

Finally, after the completion of the simulation, RadVR automatically plots the results on the corresponding  \emph{SimulationPlane} with a heatmap representation, where each sensor is located at the center of colored matrices (see Fig \ref{fig:visualization}). RadVR implements a three-color gradient palette where blue (RGB 0, 0, 255) is considered as the minimum value, yellow (RGB 255, 255, 0) as the middle value, and red (RGB 255, 0, 0) as the maximum value. For point-in-time illuminance simulations, the system automatically extract the minimum and maximum values of the simulation results, whereas in the Daylight Factor simulations the default range goes from 0 and 10. The user can later modify the minimum and maximum bounds of the visualization by accessing the corresponding range-slider in RadVR simulation menu.

\section{Case Study Applications}\label{sec:user}

To capture additional user feedback and assess how daylighting analysis tasks in RadVR can be executed, we apply the proposed tool in two case studies with two separate user groups. These studies would provide anecdotal evidence about the effectiveness of using the tool in building design and analysis.The first group, provided ongoing empirical feedback throughout the development of the RadVR software. Features such as the dynamic sunpath, locomotion, and the simulation experience were explored, with design feedback and usability testing of each of the developed functions. The second group was involved in a one-time study in comparing RadVR with a conventional 2D display daylight analysis tool, Diva for Rhino. In this anecdotal study, we explored how RadVR helped users understand the relationship between the sun, time, and the building in a sunlight study, in addition to perceiving the simulation results. Below we describe additional details of the two anecdotal studies:

\subsection{Empirical Testing}
The goal of these studies was to receive ongoing user feedback during the development of RadVR. Feature and usability testing, along with design discussions were conducted during the course of 15 weeks. Eleven students of a graduate level course - ARCH 249: Physical, Digital, Virtual- at UC Berkeley's Department of Architecture participated in this testing. The course itself was oriented towards designing interactive modules for virtual reality experiences. During weekly meetings, students tested new features of RadVR and provided empirical feedback on the developed functionalities. Features such as the direct sunlight position analysis, 9-point matrix, and locomotive modules which involved many UI elements were  discussed during these sessions. Students of this group did not necessarily come with design knowledge in daylighting, and therefore, mainly commented on the usability aspects of the virtual reality experience itself. The group provided ideas and ways of improvement for the studied features. The results of the discussion were gradually embedded into the final user experience design of RadVR. 

\subsection{Preliminary Comparative Study}
To evaluate the potential of our approach, we conducted a pilot study of RadVR with students who had previously used a popular Radiance and DAYSIM frontend for Rhino, DIVA \cite{jakubiec2011diva}. This study used the design work produced by architecture students for a daylighting assignment conducted in a graduate level course - ARCH 240: Advanced Topics in Energy and Environment - at UC Berkeley's Department of Architecture. Participation in the study was optional for students, and 16 out of 40 students volunteered in this case study. Only 5 of the 16 participants had experienced a 6DOF virtual reality experience before. The study entailed the following  phases: (1) using DIVA, improve the daylight performance of the student’s design previously done for ARCH 240 class; (2) conducting daylight analysis in RadVR with the same purpose of improving daylight performance; (3) completing an exit survey that compares the two approaches to daylighting analysis and design. 

During their previous Arch 240 assignment, students had been asked to design a 25m x 40m swimming pool facility in San Francisco, California with a variable building height. The goal of the design was to achieve a coherent and well-defined daylight concept for the building that addresses both the diffuse and direct component of light. ARCH 240 instructors advised students to consider relevant daylight strategies, including top lighting, side lighting, view out, relation with solar gains and  borrowed light. Students used the Diva for Rhino tool to assess and refine the daylighting strategies implemented in the design task phase. Daylight Factor analysis and 9 point-in-time matrix visualizations had been conducted in this phase and reported as part of the deliverables assignment. The students positioned a equally spaced analysis sensor grid, with sensors placed 0.6m from each other, at 0.8 m from the ground floor. Radiance’s ambient bounce (-aa) parameter was equal to 6. Using RadVR Grasshopper plugin (Figure \ref{fig:ghComp}), the Rhino models were exported to RadVR. The added materials followed the materials chosen by the students.

Students conducted daylight analysis in RadVR performing two tasks during a 15 minute session. Firstly, students studied the relationship between the sun, time, and the building in a sunlight study. For this, they initially navigated and inspected their designs using two implemented locomotion functions, teleportation and flying. In order to observe time variations on direct light patterns, students used the time controllers that control sun position according to hour-of-the-year and location of the design (i.e., latitude and longitude). The students also used the 9 point-in time matrix RadVR module to study the direct light patterns in different seasons and different times of day.

Secondly, the students simulated Daylight Factor through the RadVR menu. As in DIVA, they positioned the sensor grid at 0.8 meter from the ground floor. However, to reduce computation time and provide a quicker performance feedback, they used 1 X 1 m sensor grid and 2 ambient bounces. After the simulation, the participants navigated through the results, and evaluated the building design by correlating key design features that affect simulation results. The users could color mapping the results and control the value gradient by accessing visualization menus and changing the range to their preferred domain. Before each task the users followed a brief tutorial on how to use the RadVR that took approximately 4 minutes.

\thispagestyle{empty}
Upon completion, the users filled out an exit survey that evaluated their experience in RadVR when comparing to Diva for Rhino. The questions of the survey were designed  to identify tasks where the immersive experience in RadVR can bring an additional insight or improve the usability of digital simulation tools in current daylighting design workflows. The survey  had three parts. The first part evaluates the user experience in conducting the first task that focus on study direct daylight patterns overtime. The second one, focused in using RadVR for illuminance-based simulations and assessed the user experience in producing and navigating through simulation results in VR. The last part consisted in an overall evaluation (e.g., comfort, learning curve) of the RadVR software compared to DIVA for Rhino. Each question covered a specific activity of the two daylight analysis tasks. To answer to each question the students needed to choose a value of a linear 5-point scale on how they compared the performance of the mentioned activity between the two tools. To avoid any confusion, the words \textit{RadVR} and \textit{Diva for Rhino} were colored in different colors and comparison adjectives (significantly, slightly, same) were displayed as bold text. 

\subsection{Equipment}

When working with RadVR, participants used a Oculus Rift head-mounted display and controllers that was connected to a computer with a Corei7 2.40GHz processor and an NVIDIA GeForce GTX 1060 graphics card. The maximum measured luminance of the Oculus Rift display has been reported to be 80 cd/m$^2$ for a white scene RGB (255, 255, 255) \cite{Rockcastle_Chamilothori_Andersen_2017}. The Oculus Rift holds a 110° field of view display, with two OLED displays for each eye with a resolution of 1080×1200 pixels and a refresh rate of 90 Hz.

\thispagestyle{empty}
\section{Results}

Table \ref{fig:userTable}.a presents the first part of the survey results, focusing on understanding the relationship between time, the sun and the building. On average, the majority of the responses show that RadVR can be potentially helpful in understanding the variation of direct light patterns through time. As time navigation in RadVR is achieved using the VR controller joysticks, the direct sunlight penetration smoothly updates in the scene, allowing a quick understanding of how the variation of the hour and date impacts the sun location relative to the building. However, some participants pointed out the lack of the system’s ability to capturing diffuse lighting, which is a result of using biased rendering methods to achieve real-time visualizations.

The participants also found the ability to moving around in the building very useful, particularly to correlate design features and daylight performance. This highlights the importance of 6-degrees of freedom (6DOF) being available in the immersive design tool. Yet, we observed that some users experienced initial difficulties when using the available locomotion techniques (virtual flying and walking) to inspect parts of their designs in VR. This was the main drawback when compared with 2D GUI zoom and pan functionalities. Moreover, as mentioned in Section \ref{sec:9Point}, RadVR’s 9-point-matrix is not a grid of 9 rendered images of different times of the year but a matrix of 9 buttons that change the time of year of the surrounding environment. Thus, it is only possible to compare sunlight patterns of different hours and seasons by switching from a one-time event to another. The responses emphasize that comparison mostly takes place between two modes of daylighting conditions and heavily relies on visual memory. Nevertheless, the participants found that not being limited to a specific point-of-view and being able to easily change light conditions by control time and location are advantages of RadVR over daylighting analysis processes based on 2D still renderings.

\thispagestyle{empty}
 \begin{table*}
      \caption{Results of post-experiment surveys with the focus of (a) understanding the relationship between time, the sun and the building (b) understanding simulations, and (c) usability experiences. Participants were asked to indicate which software created a better workflow for the questioned activities on a five point scale.}~\label{fig:userTable} 
  \centering
  \includegraphics[width=1.8\columnwidth]{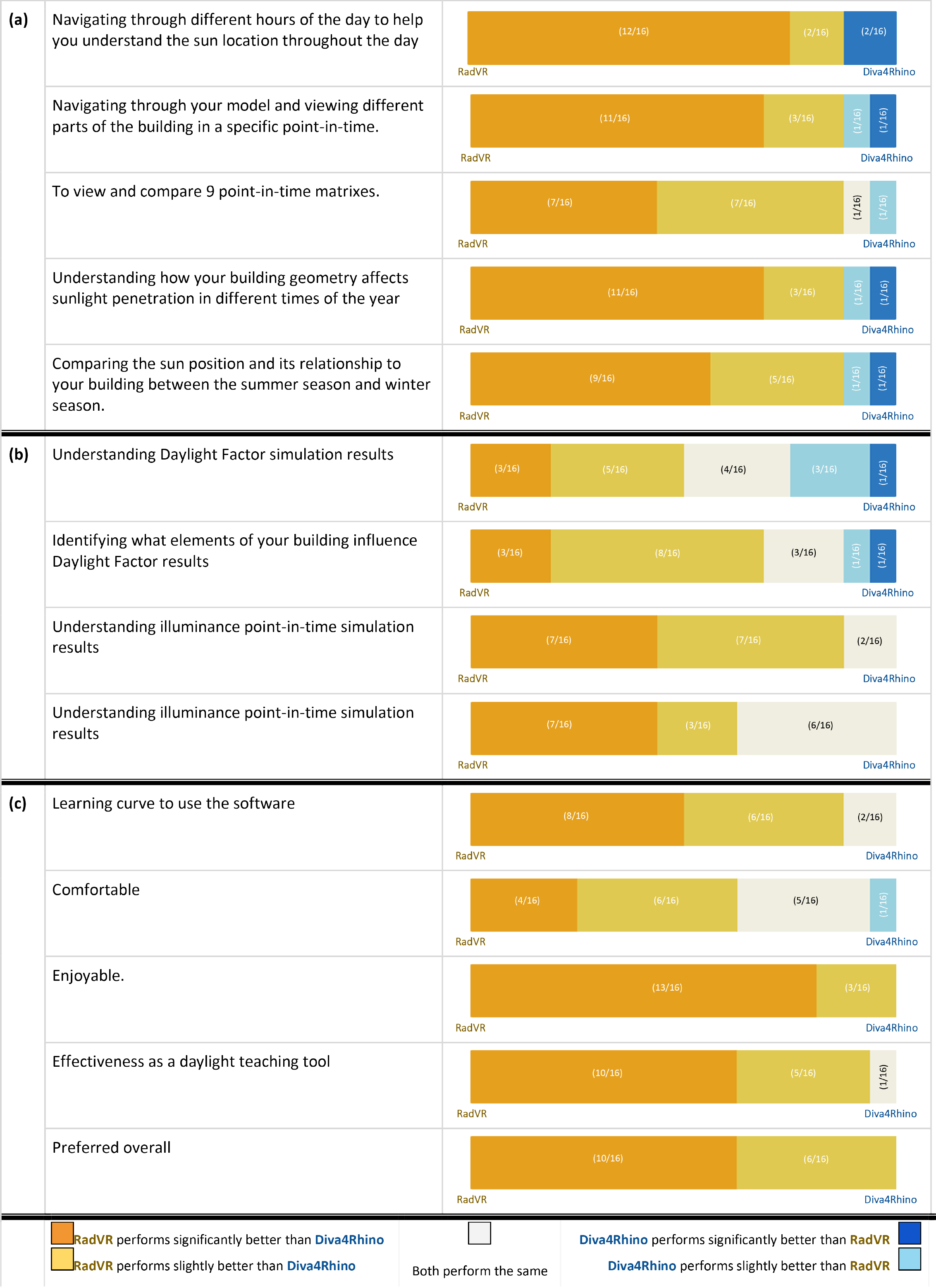}
  
\end{table*}
\thispagestyle{empty}

\begin{figure}
  \centering
  \includegraphics[width=1\columnwidth]{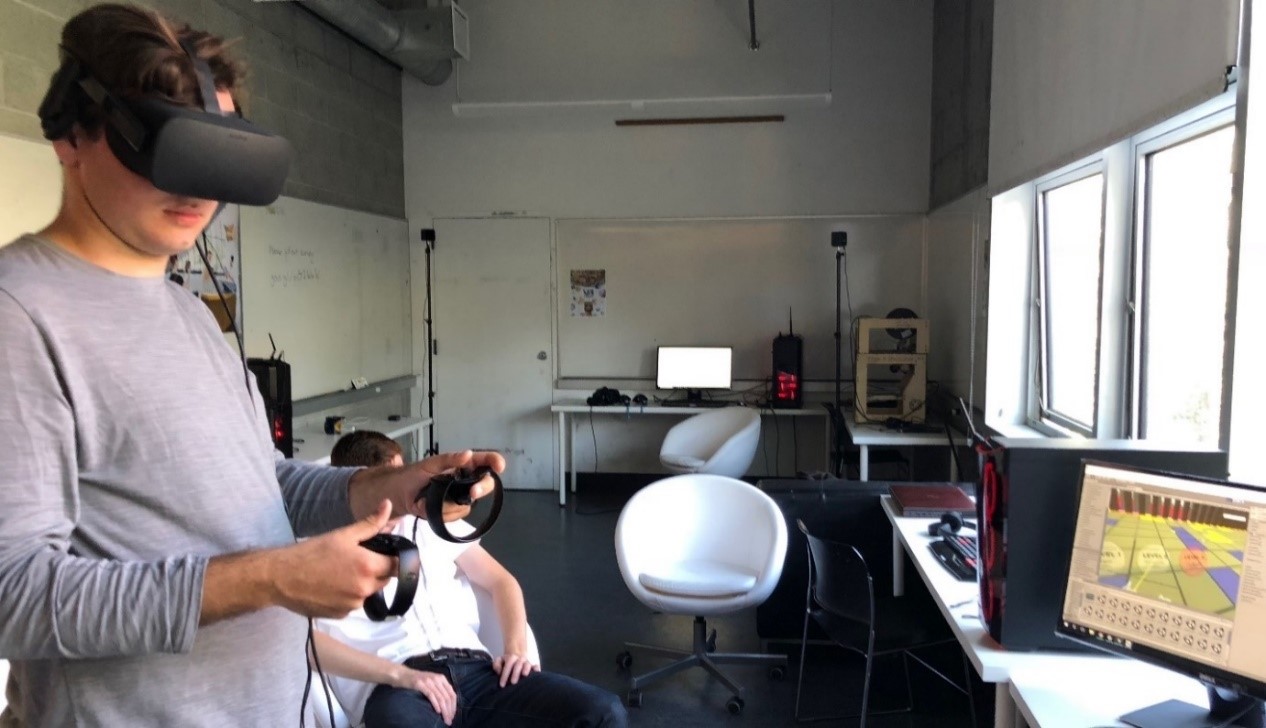}
  \caption{User experiments of RadVR while performing Daylight Factor simulations on designed spaces}~\label{fig:userPic}
\end{figure}

In the second part of the survey (Table \ref{fig:userTable}.b), the questions aimed to assess the ability of using RadVR in understanding simulation results, and their relationship with building geometry, when compared with their previous experience with DIVA. Approximately two thirds of the answers preferred RadVR as a simulation visualization tool, while the remaining indicate that the proposed system does not perform substantially better in this regard. With the  \emph{SimulationPlane} located lower than the eye level participants instantly locate over-lit or under-lit areas and virtually teleport to those areas that were outside the preferred 4-6\% daylight factor range. We also observed that students were able to locate the building elements (side openings, skylights, etc.) that affected the results, by instantly changing their point-of-view within the building and simulation map. In some cases, participants accessed the Gradient Change slider from the RadVR menu, to change the default range (0\%-10\%) of the heatmap (for example 2\%-4\%) as a way to narrow down their objective results.

Finally, in the third part of the experiment, the software’s general usability experience was studied (Table \ref{fig:userTable}.c). In terms of the learning curve, a significant majority of the participants reported that RadVR was easier to learn than DIVA for Rhino. This may be a result of designing the time navigation and locomotion functions with minimum controller interactions. However, since DIVA for Rhino was initially taught to students, many of the key concepts of daylighting analysis using software had been previously understood, and therefore, may have caused an easier learning process for RadVR. Other functions are accessed through immersive menus, which due to the large range of field of view compared to 2D screens, every window can contain many functions. During the experiment, if the subject asked on how to execute a specific function, the author would assist vocally while the subject had the headset on. 
\thispagestyle{empty}

All responses show that RadVR enabled a more enjoyable experience than DIVA for Rhino. However, since most participants (11/16) had never experienced 6DOF VR, there might be a bias towards RadVR due to its novelty and engaging environment. Many participants seemed to enjoy the new immersive experience of walking and navigating through their designed spaces with 6DOF technology, an approach which may be associated with play and in which movement is key. When asked about why RadVR could play an effective role as a teaching tool, some participants noted that the real-time update of direct light and shadows while navigating through different time events helped their understanding of the sun movement in different times of the year. After the session, a number of groups showed intent to modify their design strategies, citing their improved understanding of the geometry and its impact on daylighting performance.

\section{Conclusion}
\thispagestyle{empty}

The research proposed in this work introduces a 6DOF virtual reality tool for daylighting design and analysis, RadVR. The tool combines qualitative immersive renderings with quantitative physically correct daylighting calculations. With a user-centric design approach and an end-to-end workflow, RadVR facilitates users to (1) based on the design’s location, observe direct sunlight penetration through time by smoothly update the sun’s position, (2) interact with a 9-point-matrix of illuminance calculations for the nine most representative times of the year, (3) simulate, visualize and compare Radiance raytracing simulations of point-in-time illuminance and daylight factor directly through the system, and (4) accessing various simulations settings for different analysis strategies through the front-end virtual reality user interface.

This work includes a preliminary user-based assessment study of RadVR performance conducted with students which had previously used DIVA for Rhino. The survey results show RadVR can potentially facilitate spatial understanding tasks, navigation and sun position analysis as a complement to current 2D daylight analysis software. Additionally, participants report they can better identify what building elements impact simulation results within virtual reality. However, as the preliminary studies do not follow a fully randomized procedure with exact similar tasks and conditions for both RadVR and Diva for Rhino, the purpose is not to find if RadVR outperforms Diva for Rhino in the aforementioned tasks, but to identify tasks where RadVR can complement current two-dimensional daylight computer analysis. In fact, to fully compare the different tools a more comprehensive user study is required as future work to evaluate the effectiveness of RadVR compared with other daylighting simulation interfaces using a randomized population of users with various skill levels in daylighting design tools

Since students had initially learned the concepts of daylighting analysis through DIVA, the comparative results do not necessarily guarantee RadVR could be a substitute to DIVA, or other 2D display software, as a standalone daylighting analysis tool. Instead, we believe our proposed system indicates new directions for daylighting simulation interfaces for building design, provides additional usability, proposes new analytical procedures, and complements current daylighting analysis workflows.

\thispagestyle{empty}

One of the main contributions of this work is establishing a stable bi-directional data pipeline between Unity3D and other third-party building performance simulations tools. While many building performance simulations engines do not have native Graphical User Interfaces and can only be accessed through console-based systems, the development of a virtual reality GUI would allow architects and other building designers to conduct pre-construction analysis in 1:1 scale immersive environments of various performance metrics. Moreover, with the integration of VR design methodologies in CAD-based software, it will be possible to use such an analysis in earlier stages of design, all within immersive environments and without the need of transferring models back and forth between platforms.

However, despite the spatial immersion and presence provided by RadVR, the tool comes with a number of limitations. For example, due to the limited power of its real-time rendering system, many spatial qualities of daylit spaces cannot be captured, resulting in flat renderings and unrealistic qualitative outputs. Such limitation is mostly seen when indirect lighting strategies are implemented since biased rendering methods used in current game engine real-time rendering systems cannot fully capture ray bouncing and light scattering effects. Additionally, reading results in large scale heatmap from a human scale point-of-view is difficult, particularly with the visualized work plane usually set at 0.8 m and the eye height around 1.7 m. Users reported this limitation was rather resolved in flying mode since they could observe results from a birds-eye view. Repositioning to the right point-of-view is another limitation since it is time consuming compared to the fast orbit interactions of 3D modeling environments in 2D user interfaces. Nevertheless, after identifying over-lit or under-lit areas, users were able to teleport to the exact location and investigate what architectural element is responsible for them.

Future work on the development of this tool encompasses four main tasks. First, improve the graphics quality by implementing state-of-the-art rendering shaders and recent GPU based techniques to improve the ability of the system to better capture ambient lighting in real-time. Second, expand RadVR's current sky model palette (the CIE overcast sky and the CIE clear sky) by including all-weather Perez sky models \cite{perez1993all} which use Typical Meteorological Year hourly data to express the typical sky conditions of a given location at any given hour. The inclusion of all-weather Perez skies is also an important task to expand the RadVR tool to handle climate-based daylight metrics, including Daylight Autonomy and Useful Daylight Illuminance. Third, improving data visualizations by exploring other types of data representation that better suit 3D immersive spaces. Such an approach can enhance data reasoning tasks in VR and it can benefit from some visual properties such as stereoscopic depth or gaze and color maps. We also intend to conduct a comprehensive comparison between the RadVR sun path visualizer, other digital sun path tools, and physical heliodons to further refine RadVR sun path tool. Finally, we plan to extend current RadVR locomotion abilities with redirected walking and tunneling techniques to reduce potential motion sickness resulted from artificial locomotion in VR.

\bibliographystyle{ACM-Reference-Format}
\bibliography{sample-base}

\thispagestyle{empty}

\end{document}